\documentclass[manuscript]{aastex}
\usepackage{natbib}

\shorttitle{Post-Flare Giant Arches in the 14~Oct.~2014 Solar Eruption}
\shortauthors{M.J.~West and D.B.~Seaton}

\begin{document}

\title{SWAP Observations of Post-flare Giant Arches in the Long-Duration 14~October~2014 Solar Eruption}

\author{Matthew J. West and Daniel B. Seaton}
\email{mwest@oma.be; dseaton@oma.be}
\affil{Solar-Terrestrial Center of Excellence, SIDC, Royal Observatory of Belgium, Avenue Circulaire 3, 1180 Brussels, Belgium.}

\begin{abstract}
On 14~October~2014 the Sun Watcher with Active Pixels and Image Processing (SWAP) EUV solar telescope on-board the Project for On-Board Autonomy 2 (PROBA2) spacecraft observed an eruption that led to the formation of perhaps the largest post-eruptive loop system seen in the solar corona in solar cycle 24. The initial eruption occurred at about 18:30~UT on 14~October, behind the East Solar limb, and was observed as a a coronal mass ejection and an M2.2 solar flare. In the 48 hours following the eruption, the associated post-eruptive loops grew to a height of approximately $4\times10^{5}$~km ($>0.5\;\mathrm{R_{\odot}}$) at rates between 2--6~km~$\mathrm{s}^{-1}$. We conclude from our observations of this event that ordinary post-eruptive loops and so-called post-flare giant arches are fundamentally the same and are formed by the same magnetic reconnection mechanism.
\end{abstract}


\section{Introduction}
\label{sec:Introduction}

NOAA Active Region (AR) 12192, the most prolific of Solar Cycle 24 so far in terms of solar flares, rotated fully into view on 18~October~2014 but announced its presence on 14~October~2014 with a powerful and unusual eruption. The onset of the event occurred at approximately 18:30~UT---while the active region was still just behind the East Solar limb from the Earthward-perspective---and was marked by a coronal mass ejection (CME) with a velocity greater than $1300\;\mathrm{km\,s^{-1}}$. This CME was accompanied by a long-duration solar flare measured to be class M2.2 by the GOES spacecraft, but, since the base of the active region was largely obscured, the flare's true class was probably somewhat brighter. In the 48 hours that followed the onset of the event, a huge set of post-flare loops grew from the active region, eventually reaching a height of almost $4\times10^5$~km.

This unusually large loop system was observed by the \textit{Sun Watcher using Active Pixel System detector and Image Processing} (SWAP) EUV imager \citep{Seaton11, Halain13} on the \textit{Project for On-Board Autonomy 2} (PROBA2) spacecraft and by the \textit{Atmospheric Imaging Assembly} \citep[AIA;][]{Lemen2012} on the \textit{Solar Dynamics Observatory} (SDO). The loop system grew to such great heights that only SWAP, which has a large field-of-view, could observe the complete growth of the system. These loops appear to be EUV counterparts of the ``post-flare giant arches'' first described by \citet{deJager1985} in X-ray observations and subsequently discussed in detail in a series of papers by {\v S}vestka et al. (\citeyear{Svestka84, Svestka87, Svestka95, Svestka96, Svestka97, Svestka97b, Svestka98}), \citet{Farnik96} and \citet{Forbes00}. Our observations show that giant arches are not distinct from classical post-flare loops, but, in fact, are the same phenomenon, allowed to grow to large heights by a runaway magnetic reconnection process.

In his earlier papers, {\v S}vestka concluded that these giant arches could not be generated by the same magnetic reconnection process that generates classical post-flare loops because reconnection could not be sustained at such great heights. He argued that, because of the fall-off of magnetic field strength with height, the reconnection rate must decrease to a correspondingly negligible rate. However, \citet{Forbes00} showed that it is not the magnetic field strength that determines the reconnection rate, but rather the local Alfv{\' e}n speed. The Alfv{\' e}n speed can be written $v_{A}=B/\sqrt{\mu \rho},$ where $B$ is the magnetic field, $\mu$ is the magnetic permeability, and $\rho$ is the density. If the coronal density falls sufficiently rapidly with height, the decrease in magnetic field strength will be offset by the corresponding decrease in density. As a result, in many cases the Alfv{\' e}n speed may remain more or less constant, even at heights as large as $4\times10^5$~km.

In fact, solar eruptions are often seen with long-lasting, trailing UV/X-ray sources in the form of post-eruption loops. When observed on disk, such sources are seen as arcades of hot, bright material. When observed close to the limb, these arcades of bright loops can be observed to ``grow'' to heights of perhaps $10^5$~km over several hours. \citet{Svestka97} observed similar loop structures that grew to much larger heights than these ordinary post-flare loops in Yohkoh X-Ray images and described them as ``slowly rising giant arches.'' They reported that giant arches appeared following eruptive flares and rose with speeds of $1.1-2.4\;\mathrm{km\,s^{-1}}$ to altitudes as large as $4\times10^5$~km ($\approx0.6\;\mathrm{R_{\odot}}$).

These bright post-eruptive loop arcades are usually interpreted as a signature of hot plasma trapped on field lines that are generated by magnetic reconnection in the post-eruption current sheet. (For a complete discussion of the current understanding of eruptive flares see \citealp{Shibata11} and \citealp{Forbes96}.) In the standard model, as eruptive material moves away from the solar surface, the magnetic field lines that connect the erupting structure to the sun not only become stretched but are forced together, creating a current sheet \citep{Bruzek64}. These stretched field lines are driven into the current sheet, an X-point forms, and the field line topology is reconfigured to a less energetic state by magnetic reconnection, which facilitates the acceleration of the outgoing eruption and results in the formation of {a series of hot loop structures just above the solar surface \citep{Forbes00}. The resulting arcade of hot loops appears to grow with time as successive field lines are reconnected on top of one another and subsequently cool into observational passbands by means of conduction, radiation and convection.

The reconnection region not only reorganizes the global field but also has more local effects. In particular, it accelerates outflows---or reconnection jets---away from the X-point. These jets have been observed to interact both with the outward-directed eruption and the low-lying post-flare loop structures that form below it \citep{Savage10}. These jets also heat the surrounding plasma through thermal conduction creating a ``thermal halo,'' a process that has been both modeled \citep{Seaton2009} and observed \citep{Reeves2011}. These thermal halos are characterized by bright, very hot ($T > 10^7$~K), diffuse regions that form above post-flare loop systems. Sometimes these regions contain dynamic structures such as Supra-Arcade Downflows (SADs), which are also interpreted to be signs of reconnection outflow \citep{McKenzie2009}. The presence of these regions is thus a key sign that reconnection is ongoing in the aftermath of any eruptive event.

In this Letter we report on our observations of the 14~October~2014 eruption event and the subsequent growth of its unique post-flare loop system; these observations clearly prove \citeauthor{Forbes00}'s theoretical conclusion was correct: The growth of giant arches is not caused by a physical process distinct from that which causes more typical post-flare loops, but rather is driven by reconnection that is sustained after the initial eruption for a period much longer than is typical for eruptive flares.

\section{Observations}
\label{sec:Observations}

\begin{figure*}
\centering
\includegraphics[width=0.99\textwidth]{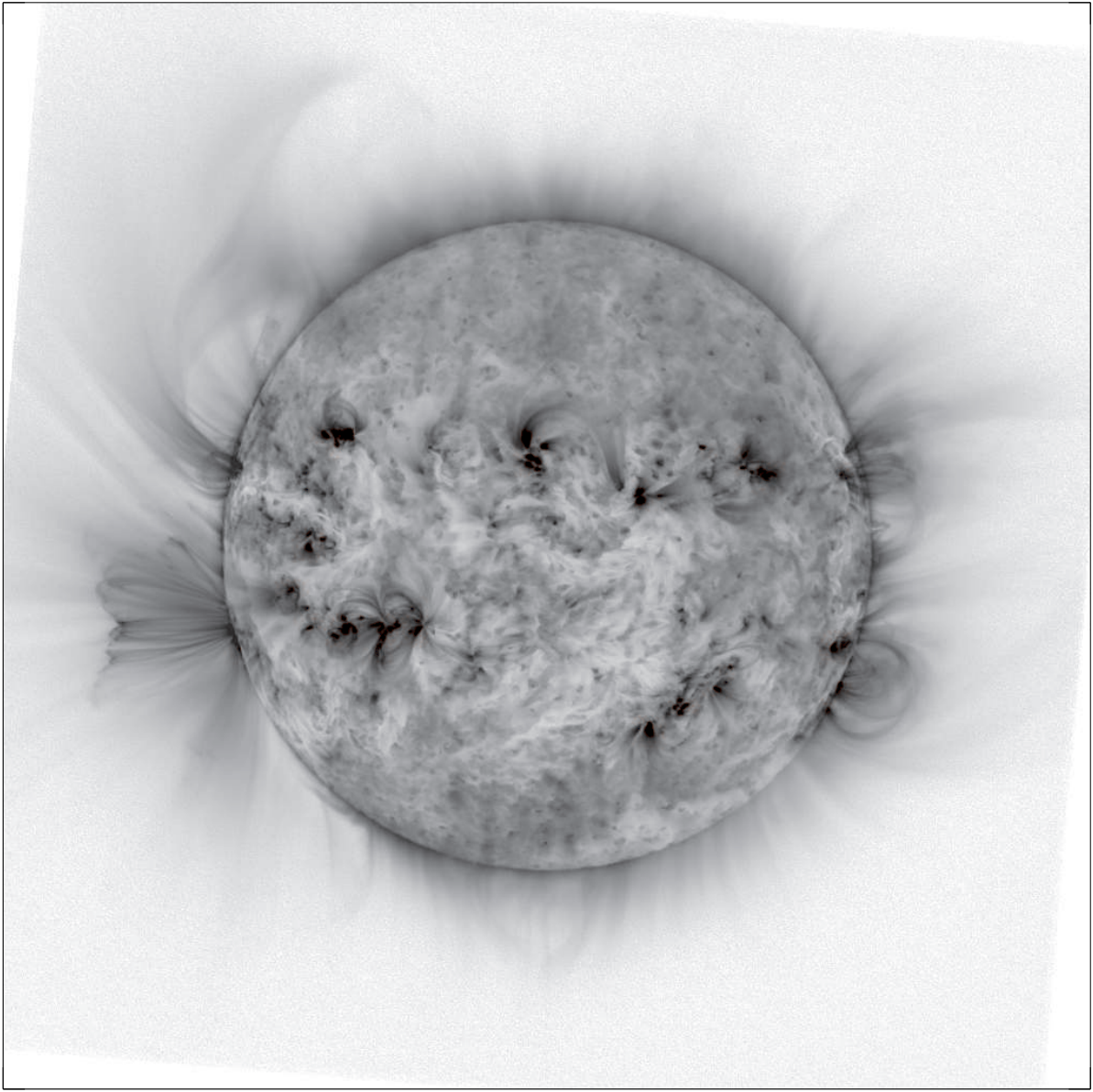}
\caption{Inverse grayscale SWAP EUV overview image on 16~October~2014 at 06:10~UT. The post-eruptive loops on the East limb can be seen almost at their maximum height at this time.}
\label{fig:Post_Arcade_Loops}
\end{figure*}

On 14~October~2014 at approximately 18:30~UT, AR 12192 produced a CME and flare just beyond the East Solar limb. Superficially, this eruption did not appear to be particularly special, but the subsequent post-flare loop system associated with it was uncommon in the duration and height of its growth. We observed this event and the post-eruptive loop system it generated using the SWAP imager between 14 and 16~October. During this period, these loops rose to a height greater than $0.5\;\mathrm{R_{\odot}}$, the largest height for any post-flare loop system observed by SWAP.

SWAP is a wide-field solar imager on-board the PROBA2 spacecraft, which orbits the Earth in a sun-synchronous dawn-dusk polar orbit. SWAP observes the Sun with a bandpass centered around 174~\AA\ (Fe~\textsc{ix/x} at $\mathrm{log\,T} \approx6$) with a cadence of roughly 100~s. SWAP images are $1024 \times 1024$ pixels in area; each pixel has a linear resolution of approximately 3.17~arcsec, producing a total field of view of approximately $54 \times 54\,\mathrm{arcmin}^{2}$. At the moment, SWAP offers the largest field of view of any EUV imager observing from the perspective of the Earth.

Figure \ref{fig:Post_Arcade_Loops} shows an inverted grayscale SWAP image of the whole Sun observed on 16~October at 06:10. The post-eruptive loops we describe here can be seen extending outwards from the East solar limb. At this time the loops were approaching their maximum height of about $0.5 \,\mathrm{R}_{\odot}$. The appearance of these loops was preceded by a long-duration M2.2 flare, observed to peak at 22:00, and a CME that was first observed at around 18:48 on 14~October. The CME had an estimated velocity of  $1360\pm561\;\mathrm{km\,s^{-1}}$ as measured by CACTUS, an automated CME tracking tool \citep{Berghmans06}. We observed an associated coronal dimming and EUV wave \citep{Moses97, Dere97, Thompson98} that emerged from the source region and traveled along the limb. STEREO-B was unavailable at this time to confirm the exact source region of the event. 

\begin{figure*}
\centering
\includegraphics[width=0.16\textwidth]{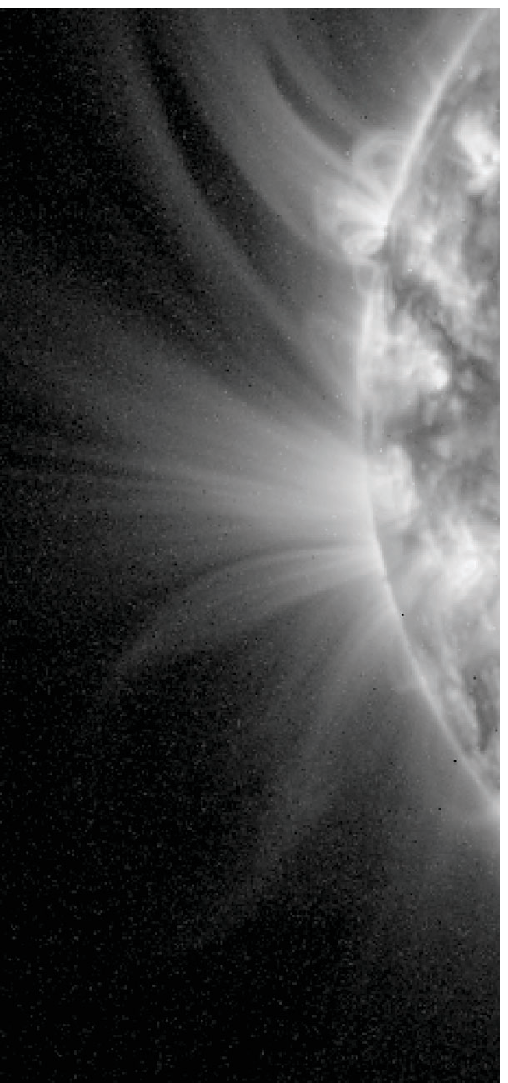}
\includegraphics[width=0.16\textwidth]{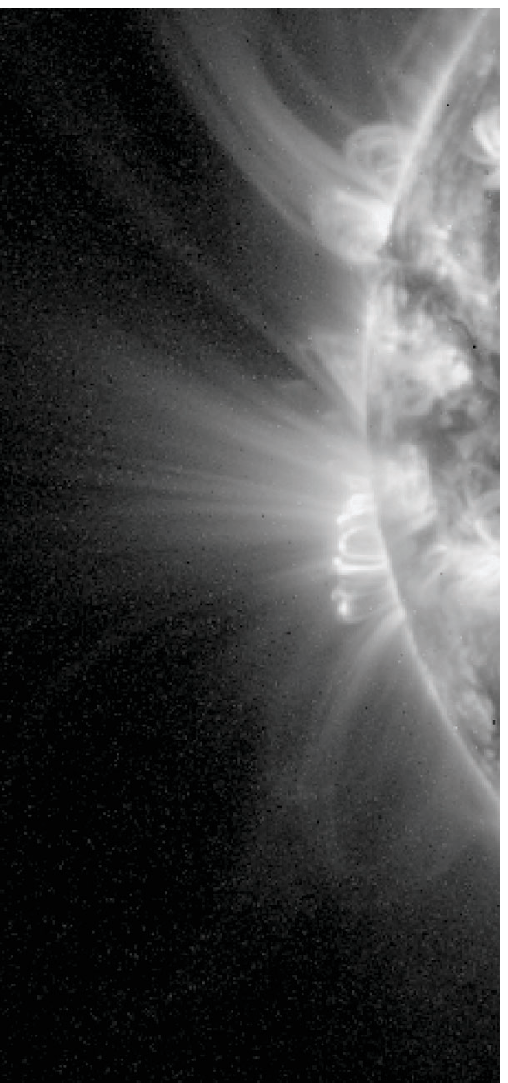}
\includegraphics[width=0.16\textwidth]{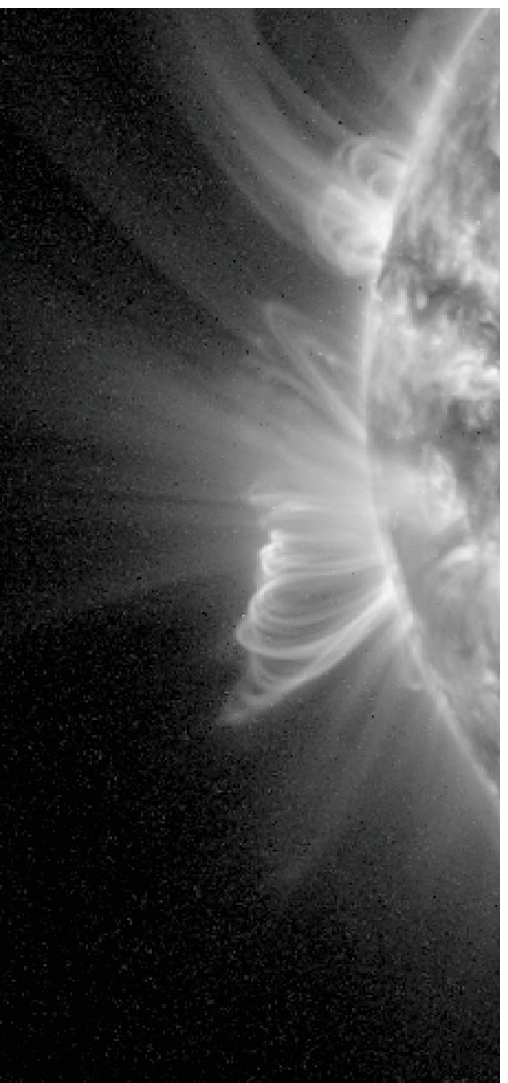}
\includegraphics[width=0.16\textwidth]{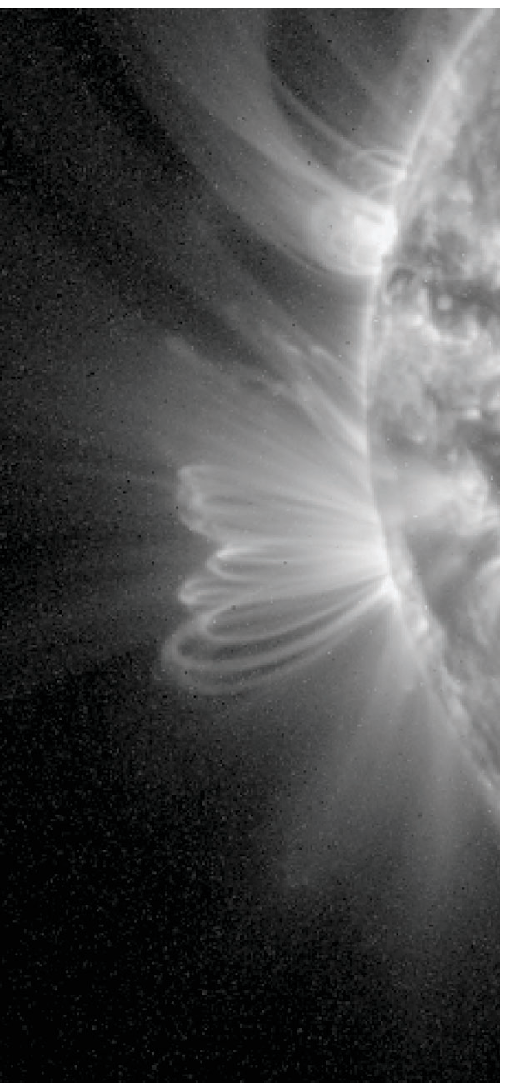}
\includegraphics[width=0.16\textwidth]{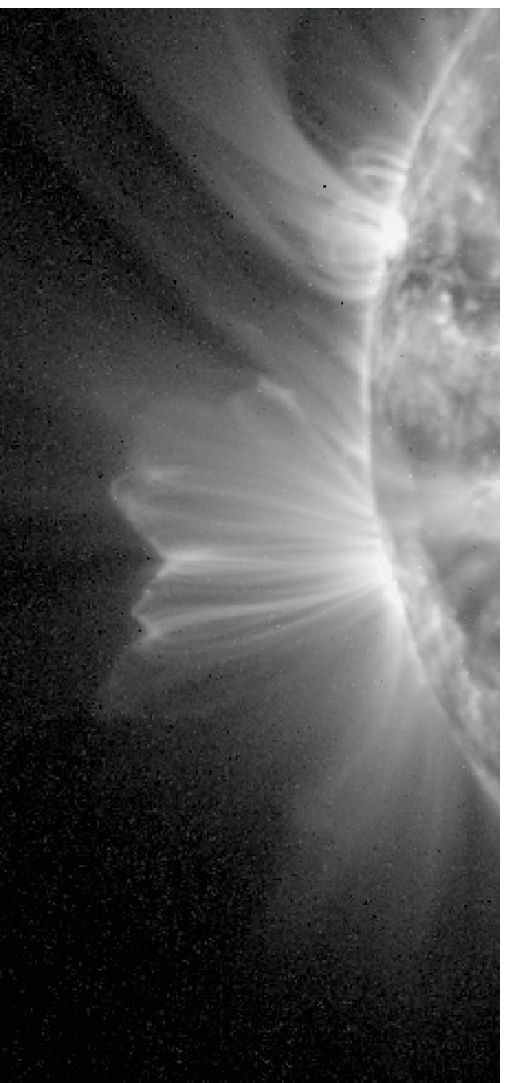}
\includegraphics[width=0.16\textwidth]{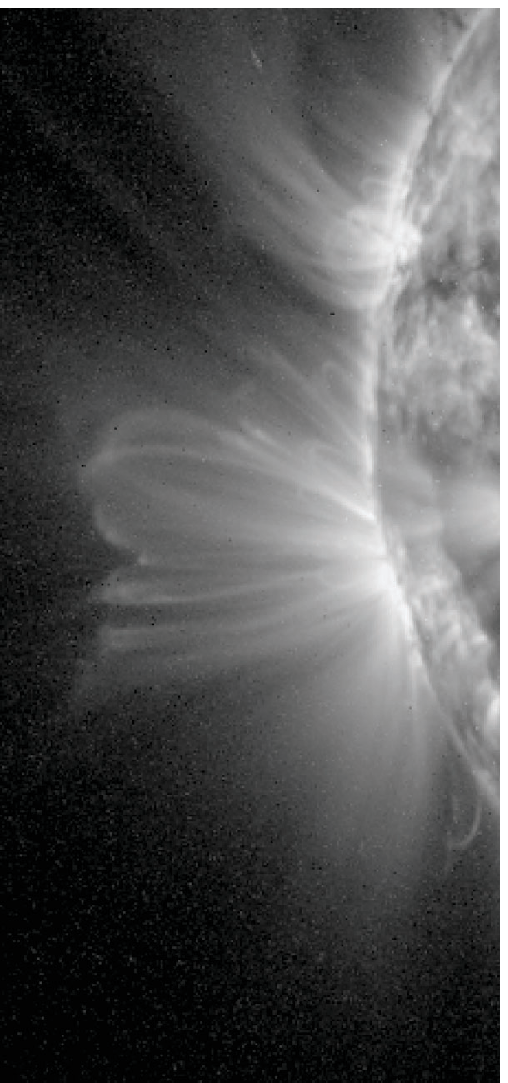}
\caption{Successive SWAP images of the emerging post-eruptive loops observed in October 2014, from left to right, at: 20:24 on 14~October; 01:15, 10:09, and 19:37 on 15~October; and 04:10 and 12:48 on 16~October. The first panel shows the pre-eruptive EUV structure, and the final image shows the system after the termination of the growth phase, when the loop structures have begun to fade. A movie is included in the online version of this Letter.}
\label{fig:loop_growth}
\end{figure*}

In the SWAP images, these post-eruptive loops are first visible at 22:55 on 14~October. The loops clearly continue to grow until about 13:00 on 16~October, after which, although they are still visible, their growth appears to be halted---or, at least, considerably slowed. Figure \ref{fig:loop_growth} shows a series of close-up images of the loops at successive times between 20:24 on 14~October and 12:48 on 16~October. The first panel shows the pre-eruptive structures on the East limb of the Sun, while the subsequent panels show the continual growth of the loop structures before they reach their maximum height. The final panel shows the loop structure after their growth has clearly ceased and the loop arcade has begun to fade.

\begin{figure*}
\centering
\includegraphics[width=0.60\textwidth]{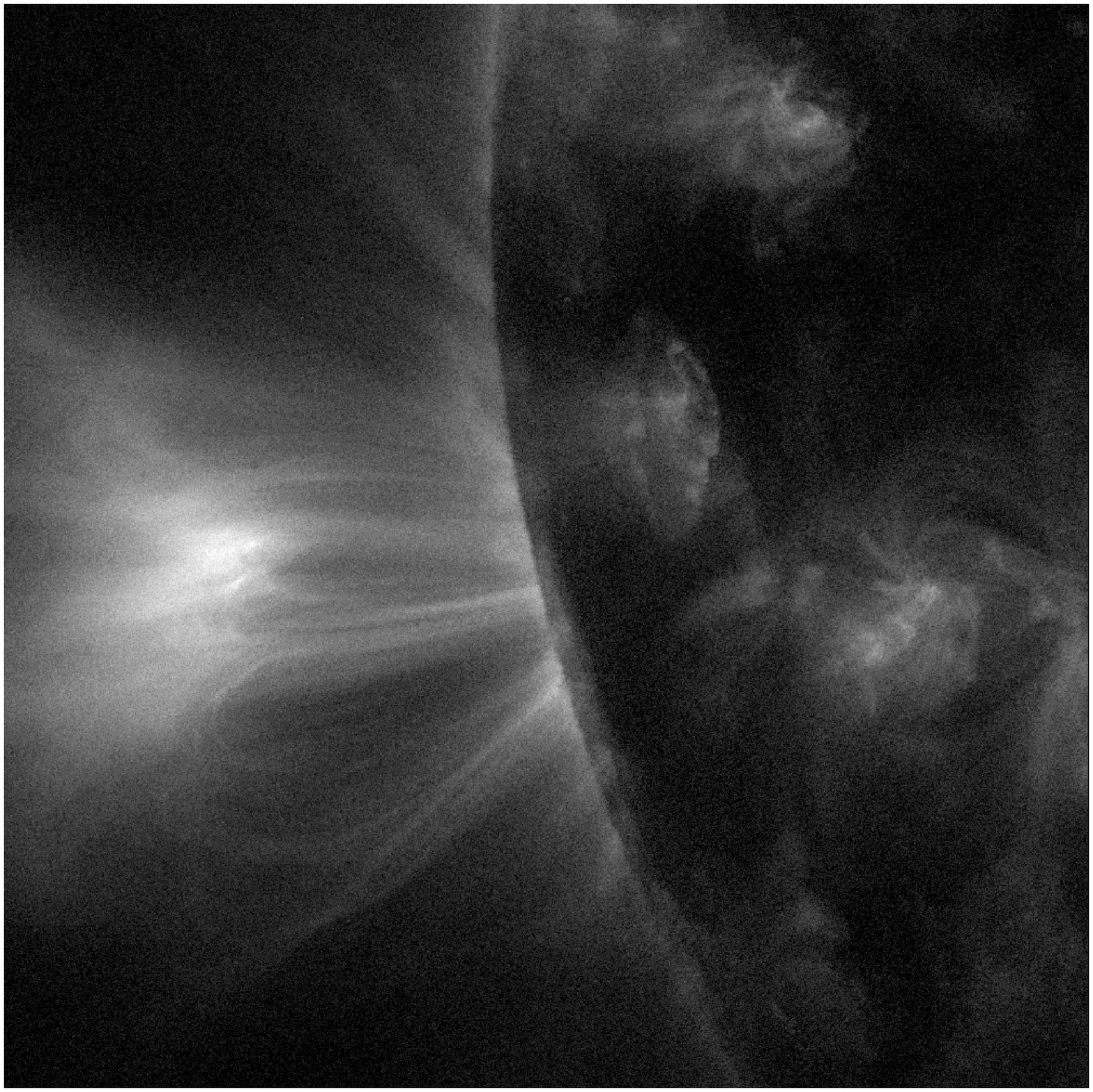}
\caption{SDO/AIA  94~\AA\ ($T=10^{6.8}$~K) image of the loop system at 10:00 on 15~October. The post-eruptive loop structures discussed in this letter are clearly visible. Above them is a bright, diffuse structure, associated with intense heating of the corona near the overlying current layer.}
\label{fig:SDO}
\end{figure*}

There is wide consensus that most post-eruptive loops are observational signatures of magnetic reconnection, whereby the reconnection of successive field lines produces a series of hot loop structures successively piled on top of each other. The loop arcade appears to grow as successive loops cool to temperatures that correspond to the production of various emission lines to which space-based EUV observatories are sensitive \citep{Hudson2011}. The reconnection process in the overlying current sheet is expected to produce plasma with temperatures in excess of $10^{7}$~K \citep{Forbes96} and, as a consequence, can heat the surrounding plasma so that it appears in EUV passbands corresponding to even higher energy transitions. Despite the loops' unusual size, our observations closely match this model of the formation of such loop systems.

Figure \ref{fig:SDO} shows an SDO AIA 94~\AA\ image of the Sun at 10:00, mid-way through the loops' growth cycle. The peak temperature response of the 94~\AA\ passband is around $T = 10^{6.8}$~K, thus the bright, diffuse emission above the post-flare loop arcade is extremely hot. Similar high-temperature features have been observed in several flares, and are interpreted to be hot halos of plasma surrounding the overlying current layer, heated by thermal conduction from the reconnecting plasma in the layer. We also see evidence of inflows, similar to the SADs described by \citet{McKenzie2009} in this region. The presence of such a halo---especially in light of the presence of SADS---is a clear indication that the reconnection process was sustained over much of the 48-hour period of growth of these loops. (Although it is worth pointing out that the growth of these loops reached such large heights that this structure was no longer present in the field of view of the high-temperature AIA passbands, and thus we cannot say with absolute certainty when it disappeared, which might more clearly indicate the point at which the reconnection process stopped.)

We measured the heights and growth rate of these loops using an automated tracking algorithm. To determine the location of the loop-tops, we took a cut through the SWAP data cube along the axis of the growing loops, in which we sampled the evolution of the brightness in each image of the data cube. This yielded a two-dimensional array in which the x-axis corresponds to time and the y-axis corresponds to height above the solar surface. For each column in the image, our code determined the loop-top location by measuring the point at which the brightness in the column falls off most rapidly with respect to a smoothed local background brightness. In general, this location corresponds with the place at which the bright loop-top falls off to the dark background above the loop. Spurious detections---which occasionally occur---were removed by eliminating points that would otherwise correspond with jumps in the measured height of the loop above a user-defined threshold. Note that, since the active region was more or less on the limb throughout the duration of the event, we neglect projection effects, which otherwise would increase the loop-top heights we measured by, at most, a few percent.

Figure~\ref{fig:growth_profile} shows a sample of the height-time array obtained for one such cut with the corresponding measured loop-top locations superimposed. As the loops dim towards the end of the event and the gradient between bright loop-top and background diminishes, our method becomes less effective. Thus there are fewer clear automated detections of the loops late in the event, even though the continued growth of the loop system is still evident to the eye.

\begin{figure*}
\centering
\includegraphics{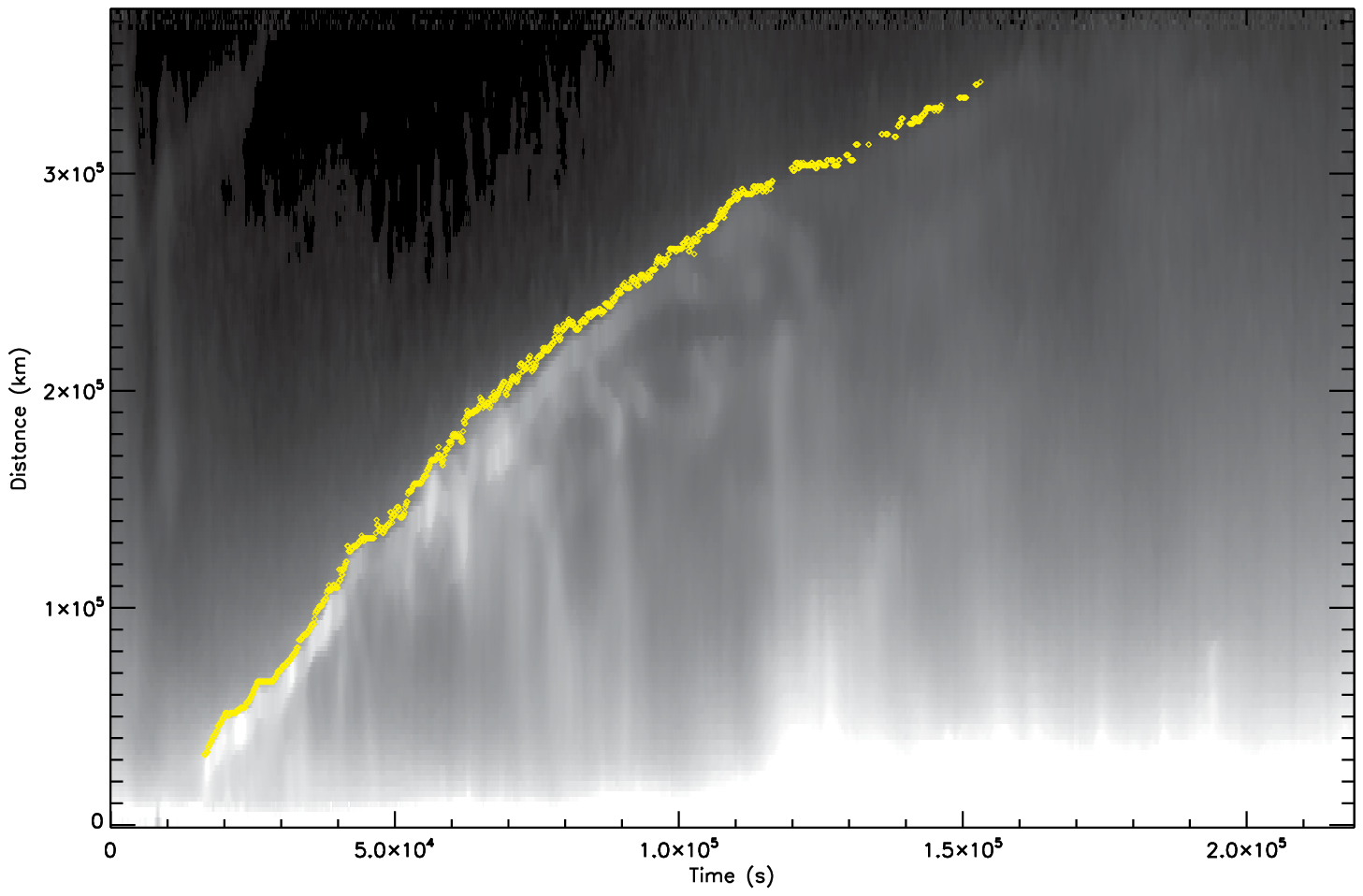}
\caption{An illustrative example of our automated loop height detection algorithm. The image shows a cut along the center axis of one loop in the system, showing its changing brightness profile over time. The superimposed yellow points show the measured height of the top of the loop.}
\label{fig:growth_profile}
\end{figure*}

We repeated this process for five different regions of the loop arcade (Figure~\ref{fig:Loop_Trajectories}), and found similar maximum heights and growth rates across the whole system. Generally speaking the loops grew at a rate of about 2--6~$\mathrm{km\,s}^{-1}$ early in the event, a rate that remained more or less steady for the first 24~hours after the eruption. The growth rate slowly decreased over the second 24~hours of our study.

\begin{figure*}
\centering
\includegraphics[width=0.39\textwidth]{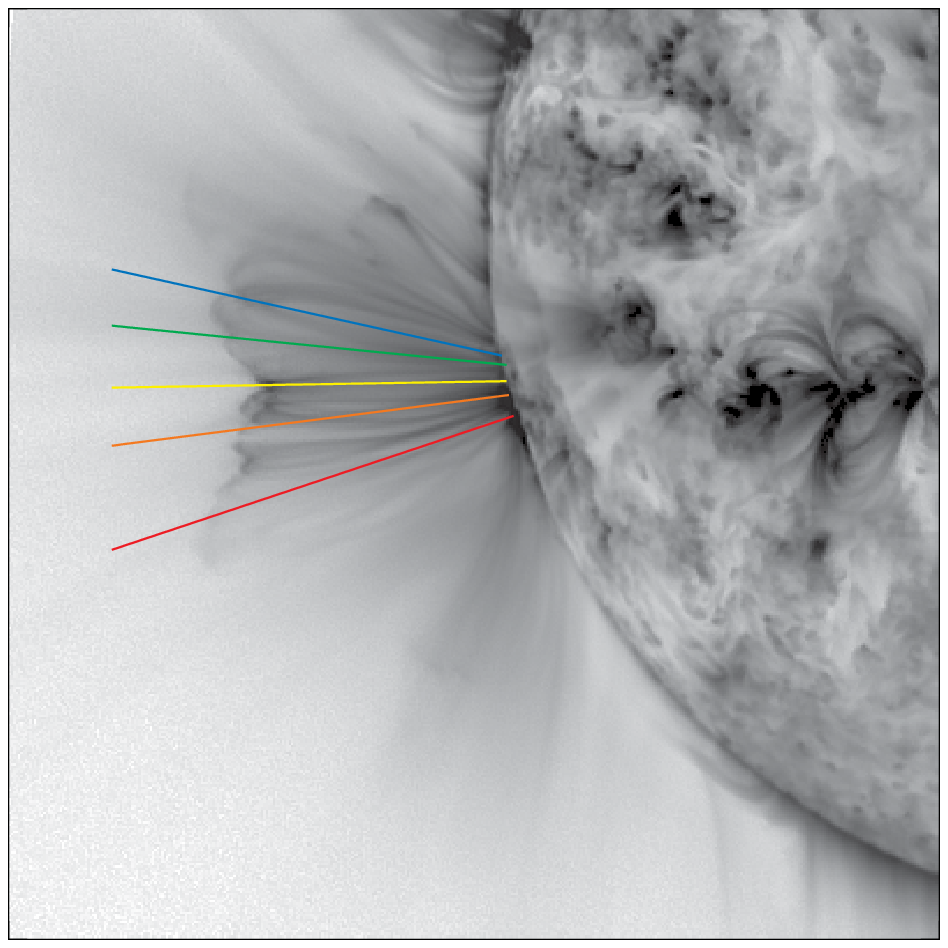}
\includegraphics[width=0.59\textwidth]{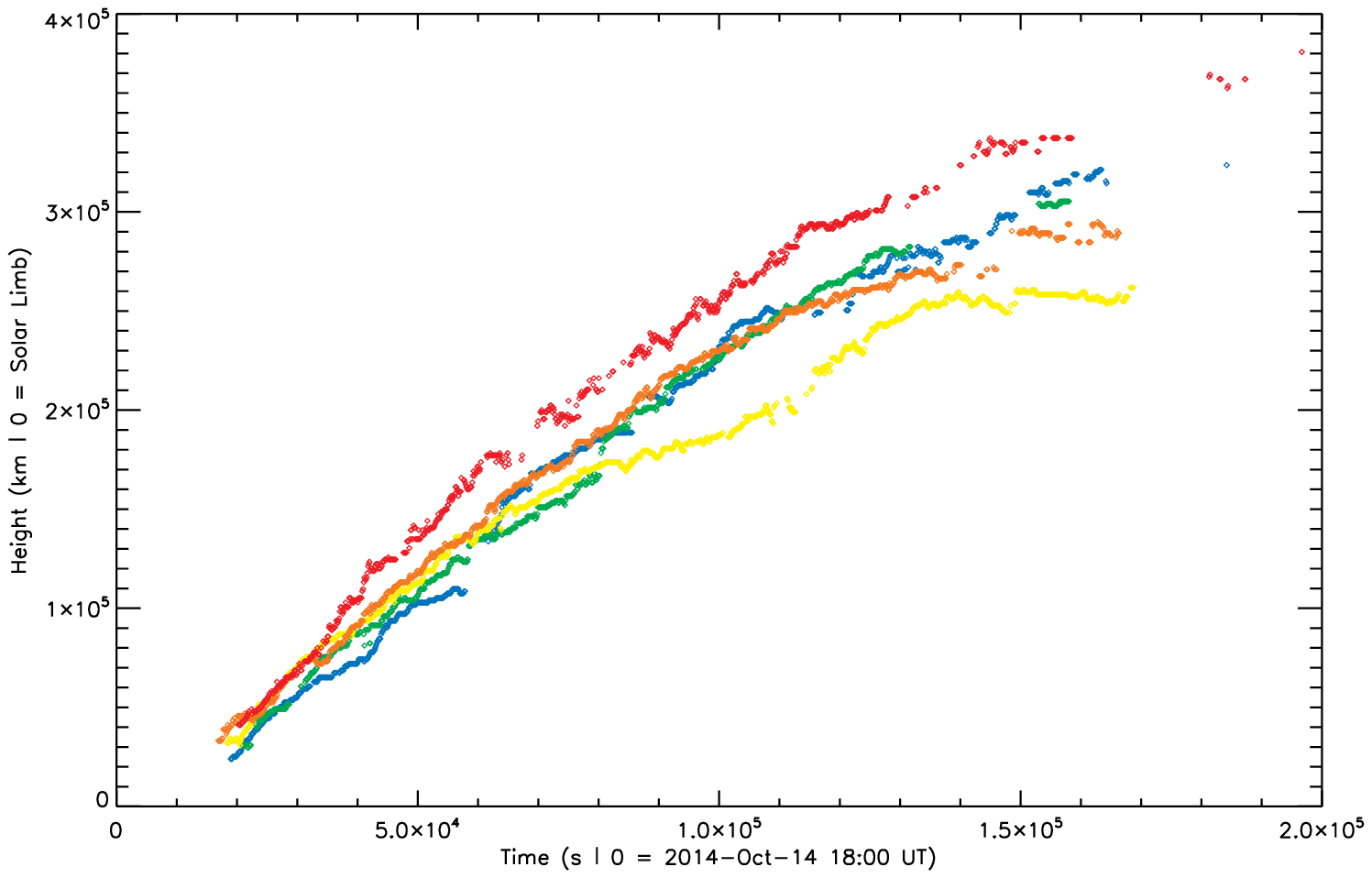}
\caption{Inverse grayscale image of the post-eruptive loops (left) observed by SWAP with several lines indicating the positions where the growth rates of loops were measured and the corresponding growth profiles in time (right). The color indicates which profile is associated with which cut in the overview image.}
\label{fig:Loop_Trajectories}
\end{figure*}

\section{Discussion and conclusions}
\label{sec:Discussion}

\citet{Svestka96} discussed observations of rising post-flare loop structures in H-$\mathrm{\alpha}$ and observations in X-rays from the Hard X-ray Imaging Spectrometer (HXIS) on-board the \textit{Solar Maximum Mission} (SMM) and the Soft X-ray Telescope (SXT) on \textit{Yohkoh}. In that paper, \citeauthor{Svestka96} distinguishes between classic post-flare loop systems, whose growth rate decreases quickly and which reach only low altitudes in the corona and so-called post-flare giant arches, which have much more steady growth rates and reach very large heights. 

\citeauthor{Svestka96} argued that the reconnection process responsible for the formation of more traditional post-eruptive loops cannot also be responsible for post-flare giant arches, because their steady growth rate requires relatively constant reconnection to be sustained for as long as 24 hours. The decreasing strength of the coronal magnetic field with height, he wrote, causes a corresponding decrease in the reconnection rate that makes such a process unsustainable.

However, \citet{Forbes00} pointed out a critical mistake in {\v S}vestka's reasoning. The reconnection rate does not depend only on the magnetic field, but rather on the local Alfv{\' e}n speed. Since the density of the corona decreases with height as well, the Alfv{\' e}n speed is---at least in some cases---relatively constant below 0.5~$\mathrm{R_{\odot}}$. Thus there is no reason that the reconnection rate would diminish substantially even in the case that the loops reach such great heights.

\citeauthor{Forbes00}'s analysis helps solve another puzzle pointed out by {\v S}vestka: in the absence of an obvious heat source, giant arches, like all loops, should cool out of X-ray and EUV passbands much more quickly than is generally observed. Thus some mechanism must work to continually heat these loops if they are to remain visible for as long as they do. What is this mechanism? In {\v S}vestka's view, either activity at the base of the loop system provides energy input, or the standard assumptions about the cooling rates due to thermal conduction in loops are incorrect. \citeauthor{Forbes00}'s analysis provides a simpler explanation: because these loops are the products of sustained magnetic reconnection, it is reconnection itself that heats the system after all.

The 14~October eruption is a clear example of the steady growth of a system of giant post-flare arches much like those described by {\v S}vestka. Furthermore, there are several signs that the reconnection process in this event was sustained for a long time. First, X-ray emission associated with the M2.2 flare caused by the eruption persisted for more than a full day after the onset of the eruption and had not yet reached the initial background level when new activity occurred, swamping the decaying signal in GOES observations. Second, bright, diffuse emission observed in the very hot 94~\AA\ passband of SDO/AIA---likely a signature of the thermal halo surrounding a reconnecting current layer---was visible until the loops became too large to be seen in AIA images, roughly 28~hours after the onset of the event.

Moreover, in the view of \citeauthor{Forbes00}, it is the steady growth of these loops themselves that is the best indication that magnetic reconnection is responsible for giant arches. Given the evidence above, we agree. This observation appears to provide the clearest evidence that giant arches and more traditional post-eruptive loop systems are fundamentally similar and are formed by the same reconnection process. However, in the case of post-flare giant arches, the reconnection process for some reason is not as self-limited as in the case of classical flare loops.

If so, these observations lead to a number of more tantalizing questions. Why doesn't the reconnection process persist for such a long time in more events? Why are {\v S}veska's so-called giant arches so uncommon? And, correspondingly, what mechanism determines when, and at what height, reconnection is switched off, limiting the growth of such systems?

One possible explanation is that the growth of many traditional post-flare loop systems really is self-limited by the Alfv{\' e}n speed in the vicinity of the erupting active region. If the density does not fall sufficiently with height, the Alfv{\' e}n speed would decrease with height and, correspondingly, the reconnection rate would drop as the height of the reconnection site increased. In that scenario the reconnection process could be self-limited, just as {\v S}vestka originally surmised. Perhaps AR~12192 is one of the unusual active regions in which the surrounding density decreases fast enough that the Alfv{\' e}n speed does not decrease significantly, allowing reconnection to proceed much longer than usual.

Indeed, there is evidence this could be the case. AR~12192 was embedded in a region of unusually EUV-bright and extended open magnetic field that persisted for at least two full solar rotations. Perhaps outflow along this field facilitated the reduction of density in the surrounding region and increased the local Alfv{\' e}n speed. If true, such an argument would suggest the need for additional investigation of the region surrounding erupting structures prior to their eruption, and possible connections both to the erupting active region and the large scale solar magnetic field.

Interestingly, in the 14 days that followed the 14~October eruption, AR~12192 produced six X-class flares and four energetic M-class flares, none of which were associated with a significant CME. Is it possible that the initial event so transformed the region that, despite its size---perhaps the largest sunspot group in over two decades---its configuration could no longer support eruptive flares? Could the sustained reconnection process that we observed have transformed so many non-potential structures overlying the region into more relaxed configurations that large scale eruptions were energetically impossible?

We cannot answer these questions in this short Letter, but it is clear that these and subsequent observations of AR~12192 could provide especially fertile ground for much research into the nature of eruptive flares and the reconnection that drives such flares.

\acknowledgments

Support for this Letter was provided by PRODEX grant No.~4000103240 managed by the European Space Agency in collaboration with the Belgian Federal Science Policy Office (BELSPO) in support of the PROBA2/SWAP mission and by the European Union's Seventh Framework Programme for Research, Technological Development and Demonstration under grant agreement No.~284461 (Project eHeroes, www.eheroes.eu). SWAP is a project of the Centre Spatial de Li\`ege and the Royal Observatory of Belgium funded by BELSPO. This paper uses data from the CACTus CME catalog, generated and maintained by the SIDC at the Royal Observatory of Belgium. We would like to thank Terry Forbes, Laurel Rachmeler and Andrei Zhukov for valuable discussions during the preparation of this paper. We also thank the anonymous referee for thoughtful suggestions that improved this Letter.

{\it Facilities:} \facility{PROBA2}, \facility{SDO}

\end{document}